\documentclass[epj]{svjour}
\usepackage{graphics}
\begin{document}
\title{Search for the Jacobi shape transition in light nuclei}

\author{A.~Maj\inst{1}, M.~Kmiecik\inst{1}, M.~Brekiesz\inst{1},
J.~Gr\c{e}bosz\inst{1}, W.~M\c{e}czy\'nski\inst{1}, J.~Stycze\'n\inst{1}, 
M.~Zi\c{e}bli\'nski\inst{1}, K.~Zuber\inst{1}, 
\\A.~Bracco\inst{2}, F.~Camera\inst{2},  G.~Benzoni\inst{2}, B.~Million\inst{2},
 N.~Blasi\inst{2}, S.~Brambilla\inst{2}, S.~Leoni\inst{2},
M.~Pignanelli\inst{2}, O.~Wieland\inst{2}, B.~Herskind\inst{3}, 
P.~Bednarczyk\inst{1,4}, D.~Curien\inst{4}, J.P.~Vivien\inst{4}\thanks{Deceased}, 
E.~Farnea\inst{5}, G.~De~Angelis\inst{6}, D.R.~Napoli\inst{6}, 
J.~Nyberg\inst{7}, M.~Kici\'nska-Habior\inst{8}, C.M.~Petrache\inst{9},
J.~Dudek\inst{4} \and K.~Pomorski\inst{10}
}                     

\offprints{Adam.Maj@ifj.edu.pl}          
\institute{The Niewodnicza\'nski Institute of Nuclear Physics,
           ul. Radzikowskiego 152, PL-31342 Krak\'ow,  Poland
 \and Dipartimento di Fisica and INFN sez. Milano, I-20133 Milano, Italy
 \and The Niels Bohr Insitute, Blegdamsvej 17, DK-2100 Copenhagen, Denmark
 \and Institut de Recherches Subatomiques, 23 rue du Loess, BP28, F-67037 Strasbourg, France
 \and INFN sez. Padova, I-35131 Padova, Italy
 \and INFN - Laboratori Nazionali di Legnaro, I-35020 Legnaro (PD), Italy
 \and Department of Radiation Sciences, Uppsala University, SE-75121 Uppsala, Sweden
 \and Institute of Experimental Physics, Warsaw University, PL-00681 Warsaw, Poland
 \and Dipartimento di Fisica, Universita di Camerino, I-62032 Camerino (MC), Italy
 \and Katedra Fizyki Teoretycznej, Uniwersytet Marii Curie-Sk\l{}odowskiej,
      PL-20031 Lublin, Poland}
\date{Received: date / Revised version: date}
%
\abstract{
The $\gamma$-rays following the reaction 105 MeV $^{18}$O+$^{28}$Si have been 
measured using the EUROBALL IV, HECTOR and EUCLIDES arrays in order to 
investigate the predicted Jacobi shape transition. The high-energy $\gamma$-ray
spectrum from the GDR decay indicates a presence of large deformations in 
hot $^{46}$Ti nucleus, in agreement with new theoretical calculations based
on the Rotating Liquid Drop model.
%
} 
\authorrunning{A. Maj et al.}
\titlerunning{Search for the Jacobi shape transition ...}
\maketitle
\section{Introduction}
\label{intro}

It is known that an atomic nucleus may change its equilibrium shape from
spherical (or prolate) to oblate with increasing angular momentum. In hot
nuclei the size of the oblate deformation increases with the angular momentum
and at certain critical value of spin an abrupt change of the equilibrium shape is 
expected, with the nucleus following a series of triaxial and more and more
elongated shapes. This phenomenon, called nuclear Jacobi
transition~\cite{Beringer} was studied in the past using, among others,
classical and semi-classical models; for a recent discussion see e.g.
Ref.~\cite{Sw_2001} and references therein. Recently developed LSD 
(Lublin-Strasbourg Drop) model~\cite{LSD} has been used to calculate
the Jacobi transition mechanism in the $^{46}$Ti nucleus; the results are
illustrated in Fig.~1.

The experimental signatures of such an abrupt change of the nuclear shape is expected
to be found in observables related to the moment of inertia.
Especially promising for the search of the Jacobi nuclear shape
transitions are: 
i)~the $\gamma$-decay of the Giant Dipole Resonance (GDR) built on such states; 
ii)~the giant backbend of the E2 $\gamma$-transition energies~\cite{Ward_2002}; 
iii)~the angular distribution of the emitted charged particles.

\begin{figure}
\resizebox{0.45\textwidth}{!}{%
  \includegraphics{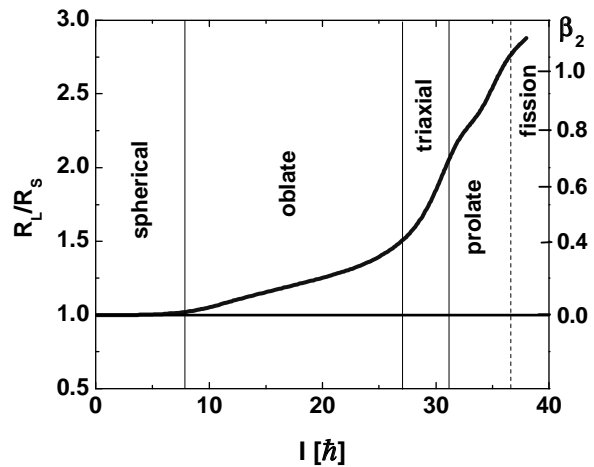}
}
\caption{Long-to-short axis ratio and the $\beta_{2}$ parameter
 for the equilibrium deformation as a function of spin, 
 obtained from the LSD model calculations for $^{46}$Ti.}
\label{fig:1}       
\end{figure}

So far, the problem of Jacobi transition in light nuclei has been addressed by
studying the GDR $\gamma$-decay from compound nuclei
$^{45}$Sc$^*$~\cite{MKH_GR93} and $^{46}$Ti$^*$~\cite{Maj_Osaka,Maj_Warszawa}.
In both cases the data indicate the presence of large deformations that could
be related to the Jacobi transition. Let us mention that in the case of
$^{45}$Sc$^*$ the GDR measurement was inclusive, while for $^{46}$Ti$^*$ it was
associated with
different $\gamma$-multiplicity values. 

In medium-mass nuclei the possible
giant backbend of the quasi-continuous E2-radiation due to the Jacobi transition
 was investigated recently by Ward et al.~\cite{Ward_2002}.

In order to address the interesting question of the Jacobi shape transition in 
more detail, and to obtain simultaneous information on the GDR decay, the E2 
quasi-continuum and charged particle angular distribution, a new experiment 
 for $^{46}$Ti$^*$
has been recently performed. In addition in this experiment,
we were studying the fission limits imposed by the angular momentum. Some results
from this measurement are presented here and discussed.

\section{The experiment}
\label{sec:1}

The experiment was performed at the VIVITRON accelerator of the IReS Laboratory of
Strasbourg (France), using the EUROBALL~IV Ge-array coupled to the HECTOR 
array~\cite{Maj94} and the charged particle detector EUCLIDES.  The $^{46}$Ti
compound nucleus was populated in the $^{18}$O+$^{28}$Si reaction  at 105~MeV
bombarding energy. The excitation energy of the $^{46}$Ti was  86~MeV  and the
maximum  angular momentum l$_{max}\approx$34$\hbar$. For this experiment
the EUROBALL consisted of 26 Germanium clover and 15 cluster detectors
(all with the BGO anti-Compton shields), and 75\% of the Inner-ball consisting
of 83 BGO crystals, which together with the  germanium detectors resulted in
the 65\% efficiency for the multiplicity determination. The 8 large volume
BaF$_2$ detectors of the HECTOR were placed in the forward  hemisphere, together
with 4 small BaF$_2$ detectors which provided a good time reference signal. The EUCLIDES
was consisting of 40 Silicon telescope detectors  covering approximately 90\%
of the solid angle. The trigger condition was such that events having at least 2
clean Ge signals  and one high-energy $\gamma$-ray were accepted. The EUCLIDES was
running in the "slave" mode, i.e. its events were accepted only if the
trigger condition was fulfilled. A total number of 10$^8$ events was collected,
in which the $\gamma$-ray energy in the Ba$F_2$ detector was
E$_{\gamma}>$4~MeV.

\section{Results and discussion}
\label{sec:2}

The results of the previous measurement~\cite{Maj_Osaka,Maj_Warszawa}
 correspond to the data gated only by
$\gamma$-ray  multiplicity; here we focus on the partial results from the new
experiment, obtained with more
restrictive condition allowing for a good identification of the  fusion-evaporation
channel.  In particular, the data shown in Fig.~2  were obtained by gating with the
known transitions in $^{42}$Ca - the residual nucleus originating 
in the decay of the  $^{46}$Ti compound nucleus at the highest angular momenta.
The left-hand side of the figure 
shows the spectrum on a logarithmic scale, together with the statistical model
calculation, in which the GDR strength function was assumed to be a 
superposition of 3 Lorentzian components. The upper right part shows the extracted
absorption cross-section  using the method described in e.g.~\cite{MKH_GR93}.
The lower right part shows the results of the shape averaging  (see
e.g.~\cite{Maj_Warszawa}) assuming a shape  distribution following the LSD
potential energy
calculations for I=22$\hbar$  (oblate equilibrium shape) and I=30$\hbar$
(elongated 3-axial equilibrium shape due to the Jacobi 
transition - see Fig.~1). As can be seen, the prediction for I=30$\hbar$
resembles
remarkably well the data, therefore suggesting the presence of the Jacobi
transition in hot $^{46}$Ti.  Further analysis, concentrating on the E2
quasi-continuum and the angular distributions of charged particles is in progress.
This is expected to provide a more consistent picture. While the GDR
$\gamma$-decay and charged particles are mainly probing the hot part of the
phase space, with the  analysis of the  E2 continuum we expect  to determine
whether or not the Jacobi shape transition is still present near  the yrast
line.

%

\begin{figure}
\resizebox{0.48\textwidth}{!}{%
  \includegraphics{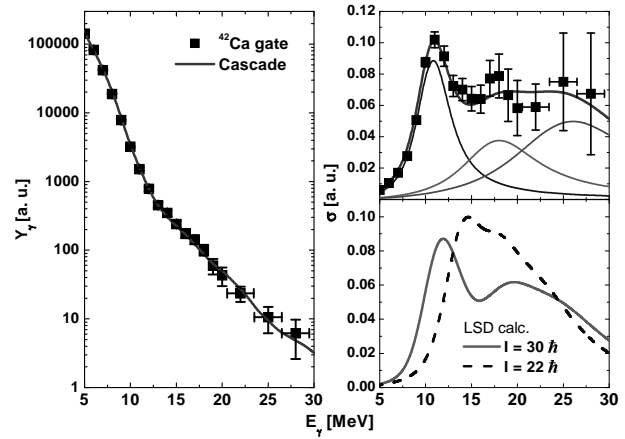}
}
\caption{Left: Spectrum of the $\gamma$-rays from the decay of the GDR
build in hot $^{46}$Ti in  coincidence with the discrete transitions 
in the residual nucleus $^{42}$Ca, together with the Cascade calculations
assuming 3 Lorentzian GDR strength function with E$_{GDR}$=10.8, 18 and 26~MeV. 
Upper right: Experimentally obtained GDR absorption cross-section and the
GDR strength function used in Cascade calculations.  
Bottom right: Thermal shape fluctuation predictions based on potential energies from
the LSD model calculations for I=22$\hbar$ and I=30$\hbar$.}
\label{fig:2}       
\end{figure}

%
%

Work supported by the Polish Committee for Scientific Research 
(KBN Grants No. 2~P03B~118~22 and No. 2~P03B~115~19), the European Commission
 contract EUROVIV, Danish Research Council and INFN.

%

%

\end{document}